\RequirePackage[preprint]{frontmatter}
\documentclass{folio}

\title{\PaperTitle}
\author{\PaperAuthors*}

\usepackage{typography}
\usepackage{mathtools}
\usepackage{citespace}
\usepackage{abbr}
\usepackage{cowriting}
\cowriter{\jie}{Jie}{olive}
\cowriter{\kexin}{Kexin}{blue}
\cowriter{\ben}{Ben}{teal}
\cowriter{\heather}{Heather}{orange}
\cowriter{\ronik}{Ronik}{violet}
\usepackage{pifont}
\usepackage{xcolor}
\usepackage[normalem]{ulem}
\usepackage{graphicx}
\usepackage{listings}
\usepackage{hyperref}
\hypersetup{colorlinks, allcolors=., linkcolor=blue!50!black, citecolor=blue!50!black}
\usepackage{entities}
\usepackage{autolabel}
\usepackage{lists}
\usepackage{comment}

\usepackage{wrapfig}
\usepackage{booktabs,colortbl,multirow,tabularx,tabulary}
\usepackage{float}
\usepackage{algorithm}
\usepackage[noEnd=false, beginLComment={//\ }, endLComment={}]{algpseudocodex}
\usepackage{tables}
\IfPackageLoadedF{subfig}{\usepackage{subcaption}}
\usepackage{tikz}
\usetikzlibrary{positioning}
\usepackage{xurl}
\usepackage[capitalize,nameinlink,noabbrev]{cleveref}

\NewDocumentEnvironment{revisedblock}{}{\ifcameraready\else\color{blue}\fi}{}

\providecommand\Description[2][]{}
\ifdefined\acks\else
\ifcameraready
\NewDocumentEnvironment{acks}{}{
  \section*{Acknowledgments}
}{}
\else
\excludecomment{acks}
\fi
\fi

\title{Natural Code Transformation Confuses \\ Agents in Vulnerability Analysis}
\begin{document}

\maketitle

\begin{abstract}

  Having demonstrated a strong potential for automating vulnerability discovery, AI agents can also be used to attack software systems at scale.
  Today's software systems need mechanisms to level the playing field by slowing down potential attacks, giving them time to patch vulnerabilities and strengthen software.
  Along this line, our study of agent trajectories in vulnerability analysis shows a surprising finding: agents produce dramatically different results across real-world code commits that carry the same functionality and the same vulnerability.
  This phenomenon appears consistently in multiple code domains and vulnerability categories.

  It motivates us to study {\em natural} and {\em functionality-preserving} code transformations derived from real-world software development, and their potential as a defense against attacker agents.
  We build a corpus of natural transformations by mining recurring development patterns from real-world code commits, and apply them to evaluate impact on agentic vulnerability analysis.
  Our results show that even a single form of transformation can effectively ``confuse'' frontier agents, e.g., Claude Mythos 5.2 with Claude Code, reducing successes in vulnerability discovery.
  We also study different agent failures, whether transformations can be composed to provide stronger protection, and adaptive attackers who seek to reverse them.
  Overall, our study of natural code transformation significantly expands the research space of using code transformations to defend against malicious agents.

\end{abstract}

\section{Introduction}

Frontier AI agents can now analyze large volumes of software code to identify vulnerabilities\citep{wang2026cybergym,lee2026sec}.
Bad actors can use them as powerful tools to discover and subsequently exploit zero-day vulnerabilities in widely used software\citep{carlini2026assessing,wang2026exploitgym,lee2026exploitbench}.
Yet many companies and institutions lack the computational resources to scan their own software using advanced agents, leaving them vulnerable to attackers wielding more powerful ones.
Less resource-rich stakeholders need a way to level the playing field, a way to slow down attackers long enough to give themselves time to identify and repair their own software.

\begin{figure}[t]
  \centering
  \begin{subcaptiongroup}
    \phantomsubcaption\label{fig:commit-stability-spend}
    \phantomsubcaption\label{fig:commit-stability-summary}
  \end{subcaptiongroup}
  \includegraphics[width=\linewidth]{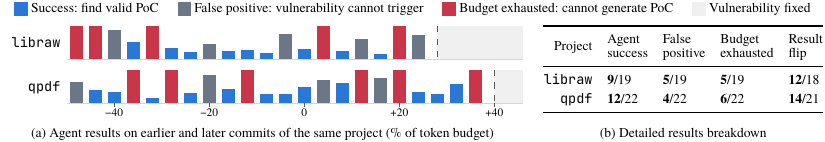}
  \caption{Vulnerability analysis results of frontier agents (GPT-5.6 Sol with Codex in this example) vary significantly across  commits in the same software.  Commit 0 is the original version evaluated by the agent;  +/- denotes preceding/following commits; the dotted line marks the vulnerability-fixing commit. Bar color = agent results; bar height = agent token cost (\% of budget). The table summarizes result variance, with ``flip'' indicating the agent's result changed between consecutive commits.}
  \label{fig:commit-stability}
\end{figure}

In our analysis of agent trajectories in  CyberGym\citep{wang2026cybergym}, we make intriguing observations that different versions of the same codebase often exhibit varying levels of resilience to AI agents scanning for vulnerabilities (\cref{subsec:commit-stability}).
More specifically, within a single repository, the agent's success rate in vulnerability analysis fluctuates significantly across adjacent code commits that carry the same functionality and the same vulnerability (\cref{fig:commit-stability}).
In many cases, the agent either fails to identify the vulnerability after exhausting the budget or generates false positives in which the proof-of-concept (PoC) input cannot trigger the vulnerability.
Such a pattern is consistently observed across multiple code domains and vulnerability categories.

This finding suggests that agents' ability to identify software vulnerabilities is highly sensitive to the code representation (i.e., syntax and structure) used in real-world software development.
It further prompts us to consider the concept of {\em natural and functionality-preserving code transformations}, and their potential as a defense to slow down or even stall attacker agents.
Conceptually, natural, functionality-preserving code transformations are like ``imperceptible adversarial perturbations'' in adversarial attacks\citep{szegedy2013intriguing}, which are intentional content modifications that preserve the original content's primary utility while manipulating how it gets interpreted by AI models.

Prior work has studied fixed code obfuscations for testing code models' robustness, applying a predefined set of local syntax-preserving code rewrites, e.g., variable renaming, equivalent control-flow rewrites like \texttt{while}-to-\texttt{for}, and unreachable-code injection~\citep{wang2023recode,mahmud2026jagged}.
However, these transformations are limited in both diversity and impact on agents (\cref{subsec:main-results}); some of them are easily identifiable and thus trivially reversible\citep{li2025systematic}.
In fact, we find prior simple transformations are not robust when the adaptive attacker agents are allowed to proactively transform the code, and they quickly become ineffective when the attacker agents are explicitly told the code is transformed so they can identify and reverse the transformations (Section~\ref{subsec:sensitivity-analysis}).

In contrast, we recognize that real-world software development activities, e.g., those exposed in commit histories, provide opportunities to identify natural code transformations that capture a much broader range of structural and semantic code changes, enabling diverse instantiations across code.
Since these transformations occur naturally during normal software development, they leave few discernible clues for agents to recognize and reverse.
These considerations make them attractive as potential techniques for code hardening against attacker agents.

To explore the feasibility of this approach, we mine recurring patterns from a large volume of real-world code commits\citep{muennighoff2024octopack}, distill them into reusable transformation skills\citep{li2026skillsbench,jiang2026sok}, and instantiate and compose them into diverse implementations based on the code context.
This leads to a large space of transformations resilient to adaptive agents attempting to reverse them. %
We validate transformed codebases to ensure that functionality and vulnerabilities are preserved modulo the ground-truth tests\citep{le2014compiler}, and evaluate their effectiveness in slowing down or stalling a variety of frontier agents. Our key findings include:

\begin{contribution}

\item \emph{Natural transformations substantially reduce frontier agents' success:}
  A single form of natural transformation can already effectively ``confuse'' frontier agents and lower their success rates significantly.
  For GPT-5.6 Sol with Codex, the agent's mean success rate falls from 63.3\% to 36.8\% ($-42\%$) under natural transformations, compared with 45.0\% ($-29\%$) under synthetic obfuscations.
  Among the agent failures, 80\% result from the agent generating a wrong PoC that cannot trigger any vulnerability, while 20\% are due to budget exhaustion.

\item \emph{Composition of transformations can be even more effective.} %
  For GPT-5.6 Sol with Codex, combining two or three transformations reduces agent success by up to an additional 40\% relative to the best single-transformation condition (\cref{tab:composition}). The agent's success rate even drops to 20.0\%, compared with 63.3\% without transformation.

\item
  \emph{Adaptive code rewriting by the agent remains ineffective.}
  Even when told which transformation was applied, the agent (GPT-5.6 Sol with Codex) recovers
  25.9 percentage points of success on synthetic obfuscations on average, but only 3.5 points on natural transformations.
  Its success on our transformed repositories remains 30.1 points below that on the original code.

\item \emph{More computation does not close the success gap within the tested limits.}
  With 4$\times$ the default budget, the agent's success on transformed repositories still trails that on the original repositories by 19.8 points, and at max reasoning effort it still trails by 25.1 points.
  The gap varies by at most 2.0 points across 1 to 9 repeated runs per task.

  \vspace{-4pt}
\end{contribution}
Our work validates the feasibility of mining real-world %
coding activities to identify rich and diverse code transformations that can protect source code against AI agents scanning for vulnerabilities.
In \cref{sec:open-challenges}, we highlight several open challenges in translating this concept into a practical and robust toolkit for code hardening against agentic vulnerability analysis.
Although related approaches have demonstrated effectiveness in traditional program analysis\citep{van2020tailoring,jung2019fuzzification,guler2019antifuzz}, extending them to agentic settings introduces new exciting open challenges.

\section{Intriguing Observations from Real-World Code Commits}
\label{sec:observations}

This work was motivated by our surprising observations from analyzing trajectories of AI agents that detect and reproduce software vulnerabilities in CyberGym\citep{wang2026cybergym}.
We discuss these findings below, beginning with a brief discussion of the problem context.

\subsection{Related Works as Context}
\label{subsec:related-work}
\vspace{-3pt}

\paragraph{Agentic vulnerability analysis}
Driven by their ability to understand and reason about code, recent work demonstrates agents' strong potential in vulnerability detection by generating proof-of-concept (PoC) input\citep{wang2026cybergym,yildiz2025benchmarking,desai2026program,zhu2026locus,luo2026agentic,zhu2026symbolon,guo2025repoaudit}, patching\citep{wang2025cve,shen2026patchbench,zhang2026bountybench,mundler2024swt}, and penetration testing\citep{shen2025pentestagent,lin2026comparing,gioacchini2025autopenbench,shao2024nyu}.
Agentic vulnerability analysis has also become the foundation for offensive tasks like exploit generation\citep{zhu2025cve, wang2026exploitgym, lee2026exploitbench, zhu2026teams,carlini2026assessing}.
Despite the strong reported performance, it remains unclear whether this performance is robust, e.g., stays consistent across variants of the same program, and can be potentially leveraged to harden the software against attackers.
This gap motivates our work.

Indeed, prior studies show that code models can be sensitive to program representations\citep{mahmud2026jagged,honarvar2026capture,thimmaiah2026llms,li2025make, hooda2025auto, pei2023exploiting}.
This raises concerns that agentic vulnerability analysis benchmarks\citep {wang2026cybergym,lee2026exploitbench,wang2026exploitgym,lee2026sec} might overlook the impact of program representation and instead attribute performance differences solely to vulnerability characteristics and agent capabilities.
Our work examines this concern by studying agent behavior under natural code transformations derived from real-world software development activities.

\paragraph{Robustness of code models}
Robustness of code models is typically evaluated using
semantics-preserving perturbations\citep{yang2024important,hossen2024adversarial,liu2025adversarial,mahmud2026jagged,henkel2022semantic,applis2021assessing,gao2023discrete} through manually crafted, context-agnostic code obfuscations, e.g., identifier renaming and dead-code insertion \citep{henkel2022semantic,mahmud2026jagged,li2025systematic}.
The local, syntactic nature of these transformations limits the diversity of program changes, often yielding recognizable, thus removable patterns (see \cref{subsec:sensitivity-analysis}).
Our work targets a different goal and considers different types of transformations, i.e., those that naturally arise during software development and are widely observed in real-world codebases.

\subsection{Agentic Vulnerability Analysis Is Susceptible to Coding Activities}
\label{subsec:commit-stability}

Real-world software development activities can change how code is organized and represented while leaving an existing vulnerability untouched.
While vulnerability analysis typically targets a single version of the software (e.g., CyberGym uses a particular commit with a known vulnerability as the ground truth), real-world software often undergoes continuous maintenance represented by a sequence of code commits.
The commits before and after the targeted version often differ only by minor changes, preserving the same functionality and often the same vulnerability.
This motivates us to ask the question: {\em can an agent's success in identifying a vulnerability on a program transfer to neighboring commits sharing the same vulnerability?}

To answer this question, we consider projects from CyberGym in which the agent has successfully identified the ground-truth vulnerability, i.e., generated a PoC input.
We then test the frontier code agent on the commits before and after this specific project version that also hold the same vulnerability (verified by replaying the PoC).
For both the original program in CyberGym and the other versions in our sampled commits, the agent is explicitly tasked with detecting the ground-truth vulnerability.

Interestingly, we find that the agent's vulnerability analysis results vary significantly across commits, even though they have been extensively pre-trained by most frontier LLMs \citep{muennighoff2024octopack}.
\cref{fig:commit-stability-spend} shows two projects included in CyberGym, while we have made consistent observations across diverse projects and LLM agents (\cref{subsec:main-results}).
Its success rate in finding the ground-truth vulnerability fluctuates, often resulting in either failing to identify the vulnerability after exhausting the budget or generating false positives where the PoC cannot trigger the vulnerability.

\begin{figure}[t]
  \centering
  \includegraphics[width=\linewidth]{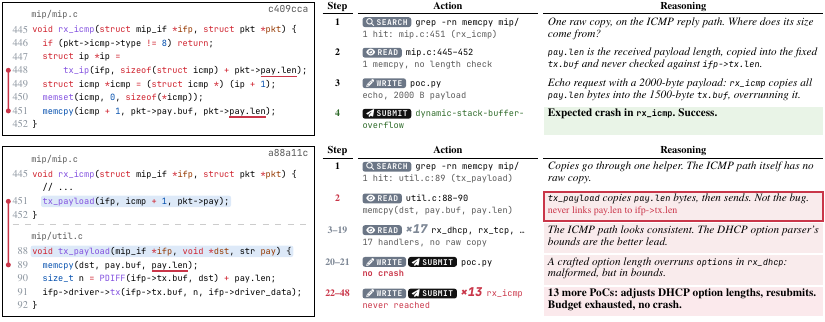}
  \caption{A small change between code commits, \texttt{c409cca} (top) and \texttt{a88a11c} (bottom), flips the agent's vulnerability analysis from success to failure. \texttt{a88a11c}'s simple code refactoring (extracting copy-and-transmit into \texttt{tx\_payload}) sends the agent to a false lead, failing to reproduce the crash.
  }
  \label{fig:motivating-example}
\end{figure}

\paragraph{A concrete example}
\cref{fig:motivating-example} illustrates how a small change in the program representation can redirect the agent's investigation.
Two adjacent commits, \texttt{c409cca} (top) and \texttt{a88a11c} (bottom), in the \texttt{Mongoose} networking library \citep{cesanta2026mongoose} contain the same vulnerability and are functionally identical.
The small code change is highlighted in \texttt{a88a11c}.
The agent locates the vulnerability in \texttt{c409cca} (reflected by its ``Action'' and ``Reasoning'' results), but fails to locate any on \texttt{a88a11c}.
The two commits differ only in that \texttt{a88a11c} places two function calls in separate files, but this separation makes it harder for the agent to link them, leading it to pursue false leads and exhaust the budget.

\paragraph{From real-world software evolution to natural code transformations}
Our findings above show that existing code development activities provide a rich set of {\em natural}, functionality-preserving code transformations that can elicit diverse agent behaviors and reasoning trajectories.
Compared to existing work studying fixed obfuscations that manipulate simple code syntax (\cref{subsec:related-work}), real-world software evolution introduces a much broader (and unpredictable) range of structural and semantic code transformations. And since these  transformations occur as part of normal code evolution, they are difficult to identify or to reverse. These properties make them attractive as potential techniques for a code hardening defense against attacker agents\citep{wang2026exploitgym, lee2026exploitbench}.

\section{Mining and Applying Natural Code Transformations}
\label{sec:approach}

Having established the potential benefits of natural code transformations, we now explore how to mine them from real-world development activities while keeping them broadly applicable to diverse codebases.
First, we mine recurring development patterns, e.g., refactoring activities, from code commits and organize them into base transformations represented as agent skills.
We then use a coding agent to localize applicable program points and implement these skills as concrete transformations.
We evaluate their efficacy as a potential defense against vulnerability analysis agents in \cref{sec:evaluation}.

\paragraph{Mining transformation categories from existing codebases}
\label{par:mining-transformations}

We group recurring development patterns into transformation categories, e.g., extracting duplicated cleanup code into a shared utility function.
The categories abstract away commit-specific names and locations so the same transformation can be applied to other repositories.
Specifically, we use CommitPackFT\citep{muennighoff2024octopack}, which contains 702,062 real-world commits across 14 major software development tasks (e.g., refactoring, cleanup, etc.) paired with instruction-like commit messages.
Given that existing agentic vulnerability analysis targets primarily memory-unsafe languages, we focus on C and C++ commits.
After removing duplicate commits and those with fewer than ten changed lines, we retain 1,196 C commits and 851 C++ commits for mining base transformations.

We iteratively apply DeepSeek-V4.1-Flash\citep{xu2026deepseek} to extract up to three atomic transformations per commit.
Each action follows the form \emph{action + concrete code element + purpose}, e.g., \emph{extract into a helper} + \emph{duplicated cleanup statements} + \emph{to centralize resource release}.
The transformation is described independently of the particular functions and variables modified by the commit.
Instructions to guide the extraction include 15 manually written seed categories as in-context examples of the desired format and level of abstraction.
Our instructions explicitly exclude identifier-only changes, added tests, and new features.
We show them in our anonymous repository.

\paragraph{From transformations to skills}
We leverage \texttt{text-embedding-3-large}\citep{openai2024embedding} to embed the transformation descriptions and apply Ward hierarchical agglomerative clustering \citep{ward1963hierarchical} to the L2-normalized embeddings.
For each cluster, we inspect representative transformations and their supporting commits, remove superficial edits such as variable renaming, and merge overlapping categories at different levels of granularity.
The resulting catalog contains 48 transformations: 25 for structure, 15 for robustness, and 8 for performance.
For evaluation purposes, we select a subset of 21 transformations that occur the most frequently in our analysis, which already cover 84\% of commits in the dataset.
As shown in \cref{fig:natural-transformations}, we represent each mined category as a skill that contains a concise instruction and a before--after code snippet \citep{li2026skillsbench,jiang2026sok}.
The instruction specifies the development action identified during mining, while the example illustrates the concrete implementation.
For example, \tf{S2} instructs the agent to extract shared logic, with an example that replaces repeated statements with calls to a helper.

\definecolor{structure-accent}{HTML}{E0A030}
\colorlet{structure}{tf-S-tint!70!structure-accent}
\colorlet{robustness}{tf-R-tint!75!tf-R}
\colorlet{performance}{tf-P-tint!75!tf-P}

\begin{figure}[t]
  \centering
  \includegraphics[width=\linewidth]{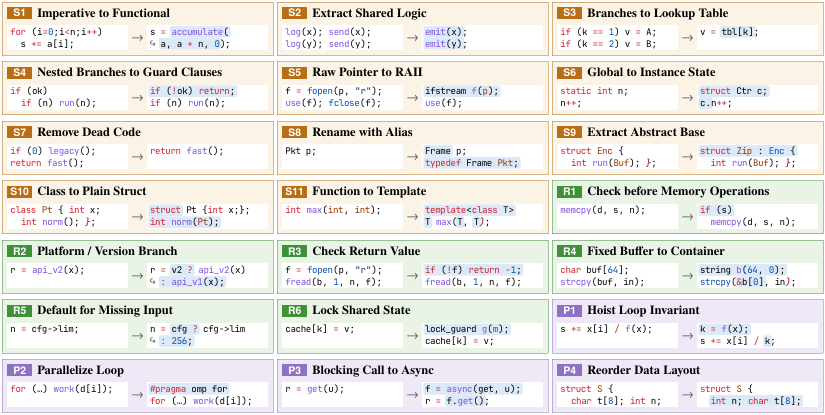}
  \caption{Top 21 most frequently occurring natural transformations identified by mining CommitPackFT, grouped into 3 categories: structure (\swatch{structure}), robustness (\swatch{robustness}), and performance (\swatch{performance}).
    Each card shows a representative code snippet  before and after the transformation. Each transformation is then represented as a skill, allowing diverse applications to target codebases.
    Unlike synthetic transformations, these transformations encode human intent and may be harder for agents to distinguish.
  }
  \label{fig:natural-transformations}
\end{figure}

Such an abstract representation of transformations enables diverse, thus less predictable applications to target codebases.
This significantly expands the search space faced by an adaptive attacker agent attempting to identify and reverse-engineer the transformations (\cref{subsec:sensitivity-analysis}), strengthening the robustness of the transformed codebase.

\paragraph{Instantiating transformations for a specific codebase}
\label{par:instantiating-transformations}

We leverage DeepSeek-V4.1-Flash to apply the mined skills to the target repository by instantiating them as concrete transformations.
Since agents have built-in stochasticity in their execution, applications of the same skill across instances can vary in the precise transformations and locations where they are applied, thus increasing the difficulty of detecting or reversing them.
We also note that transformations can be composed simply by applying transformations sequentially in diverse orders.
The specific ordering of composed transformations adds additional diversity, making reversing transformations even harder.

To ensure our transformed code matches the original in functionality, we rebuild/recompile the transformed variant and run the codebase's own test suites to ensure that the original and transformed versions are equivalent modulo the given tests \citep{le2014compiler}.
Beyond these existing tests, we apply fuzzing tests on original programs for 48 hours to collect additional input-output pairs.
We then replay these inputs on the transformed version and compare the outputs, keeping only transformations that do not introduce behavior discrepancies.
These extensive input-generation efforts give higher confidence that the transformations are semantics-preserving, even though this is unsound and does not guarantee equivalence.
We also replay the ground-truth PoC to ensure that the original vulnerability remains present.
If these tests fail, we feed the errors back to the agent as feedback to iteratively reapply the transformations.
As proving functionality-preserving can be an undecidable problem, we add a few case studies based on manual inspection of the instantiated transformations to gain more confidence by studying how they preserve the functionalities (\cref{app:manual-check}).

\section{Evaluation}

We now evaluate the impact of natural, functionality-preserving code transformations on agentic vulnerability analysis, i.e., whether/how they slow down or stall a variety of frontier agents.

\paragraph{Target code}
Our evaluation uses tasks from CyberGym\citep{wang2026cybergym}.
With CyberGym's default budget setting (\$2 or 100 steps per task), evaluating frontier agents and LLMs on all tasks across 21 transformations would require roughly \$24K in API credits for one model-harness combination.
We take the following steps to keep our costs manageable.
First, since our study focuses primarily on robustness rather than performance, we only consider tasks solvable by existing agents with sufficient budget.
Specifically, we consider tasks that GPT-5.6 Sol with Codex at extra-high reasoning successfully solves within 20 minutes.
We further excluded tasks with trivial (zero-byte) PoCs and selected a subset of the remaining samples whose distribution across PoC size, crash depth, and other features was similar to that of the original dataset.
This subset still results in approximately \$10K in API credits due to the combinations of transformed variants, models, and agents, and the default three runs per transformed sample.
CyberGym defines multiple task levels that expose different amounts of information about the target vulnerability to the agent.
In our evaluation, we use level 0 as our primary setting, where agents receive the codebase and are tasked with generating PoC input that crashes the vulnerable version and passes the patched version.

\paragraph{Agent/LLM tested}
Given the reasoning-intensive nature of our evaluation, we primarily focus on models with strong reasoning capabilities and coding performance.
We evaluate seven state-of-the-art LLMs spanning both open-weights and proprietary models: DeepSeek-V4.1-Flash\citep{xu2026deepseek}, GPT-5.6 Sol\citep{openai2026gpt56}, GPT-6 Sol\citep{openai2026gpt6}, Claude Opus 5\citep{anthropic2026claudeopus5}, Claude Sonnet 5\citep{anthropic2026claudesonnet5}, GLM-5.3\citep{glm5team2026glm5}, and Claude Mythos 5.2\citep{anthropic2026mythos} (we obtained permissions from OpenAI Daybreak and Claude Cyber Verification Program).
We pair these models with three widely used agent harnesses: Claude Code\citep{anthropic2026claudecode}, Codex\citep{openai2026codex}, and DeepSeek Harness\citep{deepseek2026harness}, based on known model-harness combinations on the CyberGym leaderboard\citep{cybergymleaderboard}.

\subsection{Impact of Code Transformations}
\label{subsec:main-results}

\begin{figure}[t]
  \centering
  \includegraphics[width=\linewidth]{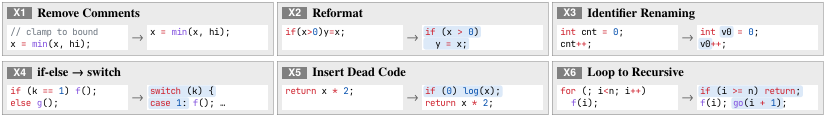}
  \caption{Six synthetic transformations widely adopted in prior robustness evaluations of code models \citep{li2025systematic,mahmud2026jagged}.
    Each card shows a representative code snippet before and after the transformation, with the specific changes highlighted.
  }
  \label{fig:synthetic-transformation}
\end{figure}

\begin{figure}[t]
  \centering
  \includegraphics[width=\linewidth]{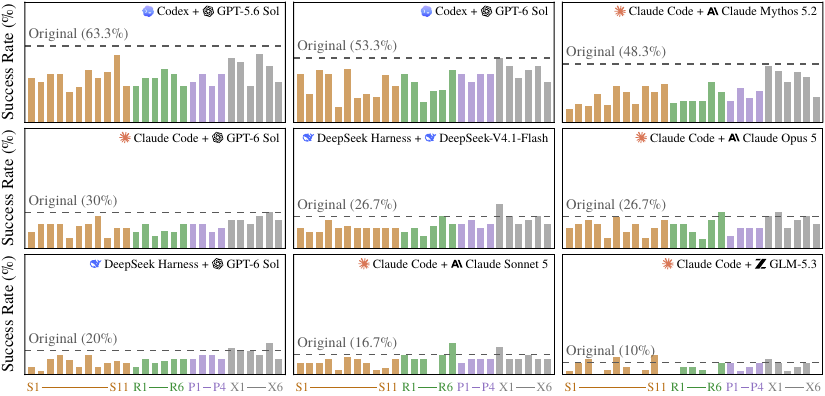}
  \caption{Agent's success rates (\%) under each natural transformation described in \cref{fig:natural-transformations} and each synthetic transformation described in \cref{fig:synthetic-transformation} for nine model--harness combinations. Dashed lines show the success rates on the original code. All panels share the same scale (y-axis). We put detailed numbers and relative changes in \cref{app:detailed-success-rates}.
  }
  \label{fig:success-rate}
\end{figure}

\cref{fig:success-rate} shows the agent's success rate across all tasks for each transformation listed in \cref{fig:natural-transformations}.
\cref{app:detailed-success-rates} provides detailed success rate values, relative changes, and comparisons across transformations and model-harness combinations.
Across the LLM agents and transformations, the mean success rate falls from 32.8\% on the original codebases to 19.1\% across transformed ones, a mean relative drop of 40.0\%.
The largest relative drop occurs for Claude Sonnet 5 with Claude Code under \tf{R4}, where the success rate drops all the way from 16.7\% to 0.0\%.

As the transformation agent is agnostic to the vulnerability (\cref{par:instantiating-transformations}), a transformation is not tailored to obscure it, and can thus occasionally make it easier to find.
We observe such gains only in three combinations of \cref{fig:success-rate}, all with Claude Code.
\tf{R6} raises the success rate of Claude Sonnet 5 from 16.7\% to 26.7\% and of Claude Opus 5 from 26.7\% to 30.0\%, and \tf{S10}, \tf{S6}, and \tf{S3} raise GLM-5.3 from 10.0\% to 16.7\%, 14.8\%, and 13.3\%, respectively.
No transformation raises the success rate of the other six combinations.
We investigate this further in \cref{subsec:sensitivity-analysis} by reversing the transformations, e.g., for \tf{S3}, \texttt{Branches to Lookup Table}, we study \texttt{Lookup Table to Branches}.

We further assess the statistical significance of these reductions using exact paired sign-flip permutation tests over the nine model-harness combinations, with 95\% confidence intervals estimated through bootstrap resampling over combinations.
All three natural transformation categories (\cref{fig:natural-transformations}) reduce the success rate of every model-harness combination by \%12.3 to \%14.1 on average ($p < 0.01$), with 95\% confidence intervals consistently excluding zero (see \cref{app:significance}).

\paragraph{Sensitivity against different harnesses}
To measure how agent performance varies across harnesses, we evaluate GPT-6 Sol with all three harnesses (\cref{fig:success-rate}).
The choice of harness substantially affects the success rate.
On the original tasks, GPT-6 Sol achieves 53.3\% with Codex, 30.0\% with Claude Code, and 20.0\% with DeepSeek Harness.
Natural transformations have a consistent relative impact, reducing these rates to average of 32.2\%, 16.7\%, and 11.5\%, respectively.

\paragraph{Comparison with synthetic transformations}
\cref{fig:success-rate} also compares natural transformations with six synthetic obfuscations extensively adopted in prior robustness evaluation of code models \citep{li2025systematic,mahmud2026jagged}.
We find that on average, these transformations decrease the agent success rate from 32.8\% to 26.9\%, whereas our transformations decrease it to 19.1\%.
Two synthetic transformations are exceptions, i.e., identifier renaming and loop-to-recursive, as we observe they are highly effective in misleading LLMs, e.g., using confusing variable names\citep{yefet2020adversarial}, but are easy to neutralize and make ineffective by adaptive attackers who also proactively transform the code, e.g., generate meaningful variable names, before the analysis (\cref{subsec:sensitivity-analysis}).

\begin{table}[t]
  \centering
  \footnotesize
  \caption{Success rates (\%) of composed transformations for GPT-5.6 Sol with Codex (left) and Claude Code (right).
  Individual success rates are shown below each composition.}
  \label{tab:composition}
  \setlength{\tabcolsep}{0pt}
  \begin{tabular*}{\linewidth}{@{\extracolsep{\fill}}l*{6}{c}@{}}
    \toprule
    & \multicolumn{4}{c}{GPT-5.6 Sol} & Opus 5 & Sonnet 5 \\
    \cmidrule[\lightrulewidth]{2-5} \cmidrule[\lightrulewidth]{6-6} \cmidrule[\lightrulewidth]{7-7}
    & \tf{S2}\,+\,\tf{R3}
    & \tf{S2}\,+\,\tf{P1}
    & \tf{S2}\,+\,\tf{R3}\,+\,\tf{P1}
    & \tf{X1}\,+\,\tf{X3}\,+\,\tf{X5}
    & \tf{R6}\,+\,\tf{S11}
    & \tf{R6}\,+\,\tf{S11} \\
    \midrule
    Composed & 26.7 & 23.3 & 20.0 & 33.3 & 16.7 & 10.0 \\
    Individual
    & 33.3\,/\,36.7
    & 33.3\,/\,33.3
    & 33.3\,/\,36.7\,/\,33.3
    & 53.3\,/\,30.0\,/\,46.7
    & 30.0\,/\,23.3
    & 26.7\,/\,13.3 \\
    \bottomrule
  \end{tabular*}
\end{table}

\paragraph{Composition}
Real-world code development often combines transformations of different kinds, so we compose transformations from different families by applying them in sequence and validating the ground-truth \poc after each step (\cref{tab:composition}).
For GPT-5.6 Sol with Codex, we compose three widely applicable transformations in each group, i.e., \tf{S2}, \tf{R3}, and \tf{P1}, into three pairs and one triple.
Every composition degrades the agent more than its most effective constituent, by 3.3 to 13.3 points.
The degradation compounds: the mean success rate falls from 34.4\% with the three constituents alone to 26.7\% with pairs and 20.0\% with the triple.
Composition can also make individually weak transformations effective.
With Claude Code, \tf{R6} alone even increases the success rate of both Claude Opus 5 and Claude Sonnet 5 above the original, and \tf{S11} alone only lowers it by 3.3 points.
However, when composed, \tf{R6}\,+\,\tf{S11} reduces the success rate of Claude Opus 5 from 26.7\% to 16.7\% and of Claude Sonnet 5 from 16.7\% to 10.0\%.

\subsection{Sensitivity Analysis}
\label{subsec:sensitivity-analysis}

We perform a series of ablations to check whether transformations consistently affect agent performance in various settings.
Similar to \cref{subsec:main-results}, we use GPT-5.6 Sol with Codex as our primary combination.
We select the two transformations most effective against this combination from each transformation family: structure (\tf{S5}, \tf{S6}), robustness (\tf{R1}, \tf{R6}), and performance (\tf{P3}, \tf{P1}).

\begin{figure}[t]
  \centering
  \begin{subcaptiongroup}
    \phantomsubcaption\label{fig:robustness-checks-budget}
    \phantomsubcaption\label{fig:robustness-checks-repeats}
    \phantomsubcaption\label{fig:robustness-checks-effort}
  \end{subcaptiongroup}
  \includegraphics[width=\linewidth]{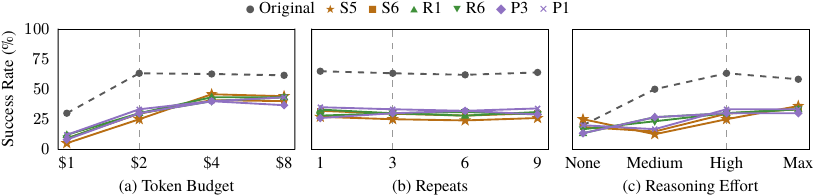}
  \caption{Sensitivity analysis for GPT-5.6 Sol with Codex on Original and the two lowest-success transformations per family: \tf{S5}/\tf{S6}, \tf{R1}/\tf{R6}, and \tf{P3}/\tf{P1}. The dashed line marks the default settings we used in \cref{subsec:main-results}. (a) Success rate against the per-task budget. (b) Mean success rate over the first $n$ of 9 runs at \$2. (c) Different reasoning effort at fixed \$2.}
  \label{fig:robustness-checks}
\end{figure}

\paragraph{Budget}
We evaluate whether changing the agent's budget impacts its success rate.
We evaluate GPT-5.6 Sol with Codex on the original codebase and under the six transformations with per-task budgets of \$1, \$2, \$4, and \$8, with the results shown in \cref{fig:robustness-checks-budget}.
Success on the original codebase stops increasing at the default \$2 budget with a success rate of 63.3\%, after which it stays consistent, but the transformed repositories keep improving up to \$4, plateauing around 41.7\% with no further improvements from increasing the budget.
While increased budgets slightly improve agent performance, they cannot restore the original success rate, even with 4$\times$ the original budget.

\paragraph{Consistency}
Since agents are inherently stochastic, we confirm whether this stochasticity impacts the consistency of the results.
Specifically, we test whether the results change depending on how many trials we run.
Using a budget of \$2, we plot the mean success rate over 1, 3, 6, and 9 repeats.
\cref{fig:robustness-checks-repeats} shows the success rate of the original task and each transformation.
We find that the results are stable regardless of how many times you repeat the test. The success rate for the original tasks stays between 62\% and 65\%, and the individual transformations do not fluctuate by more than 6 percentage points, with the success rate gap staying between 32.8 and 34.8 percentage points.

\paragraph{Reasoning effort}
Agents can reason at different effort levels, so we evaluate whether this affects the agent's success rate.
Specifically, we sweep the reasoning effort of GPT-5.6 Sol with Codex over four reasoning levels using a fixed budget of \$2.
\cref{fig:robustness-checks-effort} shows the effect of reasoning levels on the success rate.
For the original tasks, reasoning effort has a big effect, with a peak success rate at high reasoning with 63.3\%, 43.3 points higher than no reasoning, but also higher than max effort's 58.3\%.
This is expected, as reasoning improves performance, but beyond a certain point, it consumes too much of the fixed budget, degrading performance.
Interestingly, for the transformed codebase, increasing reasoning effort no longer helps as much, with the highest success at max effort being only 33.2\%.
This confirms our understanding that transformations confuse the agent, and more effort in the wrong direction will not necessarily help the agent reach the right direction.

\begin{wrapfigure}[15]{r}{0.45\linewidth}
  \centering
  \includegraphics[width=\linewidth]{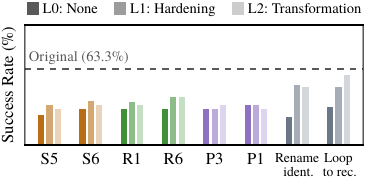}
  \caption{Success rates (\%) of GPT-5.6 Sol with Codex against an adaptive attacker. Before an attack, the agent rewrites each transformed codebase to make the vulnerability easier to find within \$8 budget. }
  \label{fig:attacker-knowledge}
\end{wrapfigure}

\paragraph{Adaptive attacker agent}
We evaluate an adaptive attacker that first rewrites the transformed codebase, then analyzes the rewritten codebase.
As shown in \cref{tab:composition}, transformations can augment each other.
We therefore study whether the attacker agent can also proactively rewrite code and reduce the effectiveness of transformations applied to resist their analysis.
Specifically, the attacker agent operates in two stages.
In the first stage, it receives the transformed codebase and a \$8 budget, 4$\times$ the analysis budget, and attempts to transform the code to make the vulnerability easier to find.
For the rewritten stage, we consider three knowledge levels: (L0) where the agent is given no information about the defense; (L1) where it is told that the codebase has been hardened using code transformations; and (L2) where it is additionally told which transformation was applied.
\cref{fig:attacker-knowledge} shows that our natural transformations remain effective at every level, likely because they are harder to detect and reverse, as they naturally arise in everyday software development activities.
In contrast, with increased knowledge of the hardening transformations, the attacker agent easily undoes the synthetic transformations shown to be highly effective in making the attacker agent fail (\cref{fig:success-rate}).

\begin{figure}[t]
  \centering
  \begin{subcaptiongroup}
    \phantomsubcaption\label{fig:ablations-direction}
    \phantomsubcaption\label{fig:ablations-implementation}
    \phantomsubcaption\label{fig:ablations-scope}
  \end{subcaptiongroup}
  \includegraphics[width=\linewidth]{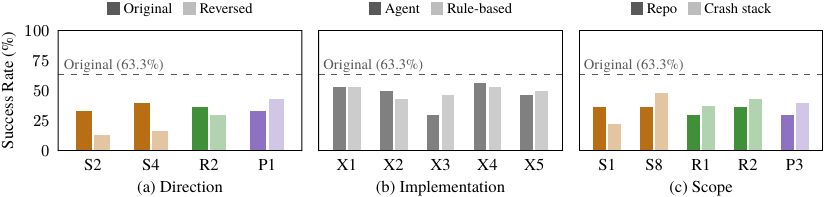}
  \caption{Success rates (\%) of GPT-5.6 Sol with Codex against transformations with (a) Original direction vs. reversed direction. (b) Implemented by agent vs. fixed rules. (c) varying scope.}
  \label{fig:ablations}
\end{figure}

\paragraph{Reversed transformations}
Each natural transformation in \cref{fig:natural-transformations} applies a development action in one direction, but developers also perform many of these actions in reverse.
To test whether the degradation depends on the direction of the edit, we reverse four transformations that span all three families and apply each reverse to the original repositories.
Specifically, reversing \tf{S2} inlines shared helpers into their call sites, reversing \tf{S4} nests guard clauses back into branches, reversing \tf{R2} removes platform or version branches, and reversing \tf{P1} moves invariants back into their loops.
\cref{fig:ablations-direction} suggests that degradation depends more on the edit itself than its direction, expanding the set of transformations available for defense.

\paragraph{Effectiveness of transformation implementation}
Our approach uses agents to apply transformations (\cref{par:instantiating-transformations}), but prior work often implements them with rule-based code rewrites, so we isolate the effectiveness of the implementation.
We use \texttt{tree-sitter}\citep{tree-sitter} to apply rule-based transformations and compare them with agent-based implementations of the five synthetic transformations \tf{X1}--\tf{X5}.
As \cref{fig:ablations-implementation} shows, both implementations obtain similar attacker success rates for four of the five transformations.
The exception is identifier renaming (\tf{X3}), where the agent generates more misleading names based on the local context, potentially confuses agent more.

\paragraph{Sensitivity to transformation scope}
To test whether degradation is simply caused by edits near the vulnerability, we compare our default setting, which allows edits anywhere in the repository, with the same four transformations restricted to functions on the crash stack.
As \cref{fig:ablations-scope} shows, restricting the transformation scope increases the success rate for four of the five transformations.
This shows that the transformation location matters, but transformation size is less important (see \cref{app:size-locality}).

\subsection{Failure Mode Analysis}

In our setting, an agent can fail either because it runs out of budget or because it generates an invalid PoC.
Having observed that natural transformations reduce agent success rates, we now study which failure modes they induce.
We compare trajectories from successful agent runs on the original tasks with those from failed agent runs on the transformed tasks.
Specifically, we examine 6 tasks in which an agent (GPT-5.6 Sol with Codex) succeeded on the original codebase but failed on most of the transformed codebase, with 6 successful trajectories and 109 failed trajectories.

\paragraph{Budget exhausted vs invalid PoC}
Under natural transformations, the agent only exhausts its budget in 20\% of failure cases.
In the other 80\% of failure cases, the agent generates a full PoC file and chooses to stop.
In these 80\% of cases, the PoC does not actually trigger a vulnerability, i.e., it is invalid or a false positive.
Another observation is that 17 out of 21 transformations have at least one trajectory that exhausts the budget, covering 4 out of 6 tasks we analyze.
This shows that exhausting the budget was not unique to any particular transformation or task.

\paragraph{Increasing budget does not help}
As shown in \cref{fig:robustness-checks-budget}, even with 4$\times$ the default budget, the agent is unable to recover to the baseline success rate.
In fact, we find failure cases tend to use \textit{fewer} tokens by giving up early, suggesting that budget was not the issue, e.g., failure cases generate 7,091 output tokens on average, while the successful ones generate 8,685 on average.

\section{Open Challenges and Future Work}
\label{sec:open-challenges}

While our work presents an intriguing finding, it warrants further investigation and optimization.
First, our work uses code agents to extract transformation patterns from existing code and to apply them to target code.
While effective, this limits interpretability, which can be significantly improved by incorporating humans in the loop.
Second, our mining effort focuses on a single dataset of 700+K commits.
We expect many more transformations can be extracted from larger C/C++ codebases.
Finally, we evaluate agents on the hardest CyberGym setting (level 0) to demonstrate the impact of transformations.
Future work should expand to a broader set of settings.

Meanwhile, our work expands the research space of using code transformations to resist
unauthorized code scanning by AI agents. Future research needs to address questions on fundamental mechanisms underlying code transformations and their efficacy against diverse agents.

\paragraph{Reversibility} Compared to fixed code transformations such as loop unrolling, code transformations extracted from repositories are significantly less vulnerable to automated identification and removal by agents. One promising direction is the exploration of one-way transformations that utilize randomization to prevent automated reversal by AI agents.
Such techniques would make transformations more robust against adaptive agent attempting to make .

\paragraph{Interpretable analysis of transformations}
Our results demonstrate agents' surprising susceptibility to natural code transformations, which should have been extensively pre-trained by code LLMs.
Deeper analysis of agent failures in response to specific transformations can shed light on fundamental limitations to agent reasoning and analysis, possibly offering new steps to better control of offensive code agents and defenses against them.

\paragraph{Agentic countermeasures}
Beyond simply telling the agent about the transformations, we need to explore more powerful countermeasures agents can apply against defensive code transformations.
The fact that frontier code LLMs remain susceptible to code commits (\cref{fig:commit-stability}) shows that data-driven approaches, e.g., pre-training with data augmentation, do not serve an effective approach.

\paragraph{Semantics-preserving transformations by construction}
We validate that our transformations preserve program functionality modulo ground-truth tests.
Stronger guarantees on preserving program semantics by construction can improve the usability and deployability of these kinds of code transformations in realistic software development settings.

\bibliography{references}

\clearpage
\appendix

\section{Data Sampling Stats}
\label{app:task-sampling}

\begin{figure}[H]
  \centering
  \includegraphics[width=\linewidth]{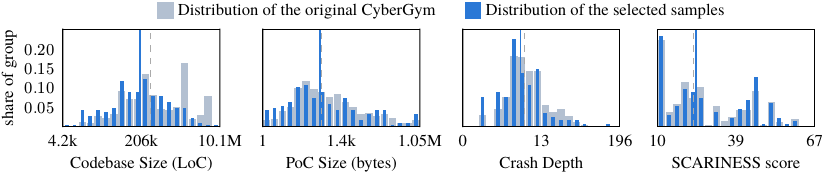}
  \caption{Feature distributions in the original CyberGym benchmark (gray) and the 186 selected tasks (blue). Bars show within-group proportions among tasks with observed values. Dashed gray and solid blue lines mark the respective medians. Codebase size, PoC size, and crash depth use $\log_{10}(x+1)$ binning, while SCARINESS scores use linear binning.}
  \label{fig:feature-histograms}
\end{figure}

We selected our tasks from the popular CyberGym benchmark \citep{wang2026cybergym}, which contains 1,507 repo-level vulnerability detection tasks with ground-truth PoCs from 188 real-world open-source projects, covering 28 crash types. We construct a smaller subset that still preserves the distribution of the original benchmark and covers diverse characteristics. Since there are 188 source projects used in CyberGym, we made sure to get exactly one task from each project. However, there were 151 tasks with zero-byte \poc{}s or repeated crash-site reports. After excluding these, we were left with 186 projects.

\paragraph{Representative subset.} We wanted our subset to be representative of the CyberGym dataset, so we focused on balancing the project source, codebase size (lines of code), the crash type, PoC size (in bytes), crash stack depth, and the SCARINESS (sanitizer) score used in CyberGym's original sampling method \citep{wang2026cybergym}. This gave us 186 tasks from the 186 projects covering 26 crash types, and the numerical feature distributions are visualized in \cref{fig:feature-histograms}. As a minor difference, our subset reduces the share of heap-buffer-overflow tasks from 38.8\% to 25.3\% and increases the share of UndefinedBehaviorSanitizer reports from 6.6\% to 13.4\%. Given that balance is useful in this kind of sampling, and there were only three sanitizers used, this is a reasonable change.

\section{Detailed Success Rates}

\definecolor{drop}{HTML}{C93649}
\definecolor{gain}{HTML}{237A3B}
\begin{table}[htbp]
  \centering
  \scriptsize
  \setlength{\tabcolsep}{0.5pt}
  \caption{Success rates (\%) under each natural transformation in \cref{fig:natural-transformations}. Each cell shows the absolute rate above its signed relative change $100(s/s_{\mathrm{Orig}}-1)$: \textcolor{drop}{negative} indicates a reduction, \textcolor{gain}{positive} an increase. Values are truncated to integers after calculation; Orig shows only its absolute rate. Dashes denote missing or undefined values. Means require all nine model--harness combinations, with relative changes computed from the mean rates.}
  \label{tab:success-rate}
  \begin{tabular*}{\linewidth}{@{\extracolsep{\fill}}l*{22}{c}@{}}
  \toprule
  \multirow{2}{*}{\textbf{Model--harness}} & \multirow{2}{*}{\textbf{Orig}} & \multicolumn{11}{c}{\cellcolor{tf-S-tint}\textbf{Structure}} & \multicolumn{6}{c}{\cellcolor{tf-R-tint}\textbf{Robustness}} & \multicolumn{4}{c}{\cellcolor{tf-P-tint}\textbf{Performance}} \\
  &  & \cellcolor{tf-S-tint}\tf{S1} & \cellcolor{tf-S-tint}\tf{S2} & \cellcolor{tf-S-tint}\tf{S3} & \cellcolor{tf-S-tint}\tf{S4} & \cellcolor{tf-S-tint}\tf{S5} & \cellcolor{tf-S-tint}\tf{S6} & \cellcolor{tf-S-tint}\tf{S7} & \cellcolor{tf-S-tint}\tf{S8} & \cellcolor{tf-S-tint}\tf{S9} & \cellcolor{tf-S-tint}\tf{S10} & \cellcolor{tf-S-tint}\tf{S11} & \cellcolor{tf-R-tint}\tf{R1} & \cellcolor{tf-R-tint}\tf{R2} & \cellcolor{tf-R-tint}\tf{R3} & \cellcolor{tf-R-tint}\tf{R4} & \cellcolor{tf-R-tint}\tf{R5} & \cellcolor{tf-R-tint}\tf{R6} & \cellcolor{tf-P-tint}\tf{P1} & \cellcolor{tf-P-tint}\tf{P2} & \cellcolor{tf-P-tint}\tf{P3} & \cellcolor{tf-P-tint}\tf{P4} \\
  \midrule
  \shortstack[l]{\model{deepseek}{DS V4.1 Flash}\\[-2pt]{\fontsize{6}{7}\selectfont\color{black!70}DeepSeek Harness}} & \shortstack[c]{26\\[-2pt]{\fontsize{6}{7}\selectfont\color{black!70}\phantom{0}}} & \shortstack[c]{16\\[-2pt]{\fontsize{6}{7}\selectfont\color{black!70}\textcolor{drop}{\textminus{}37}}} & \shortstack[c]{13\\[-2pt]{\fontsize{6}{7}\selectfont\color{black!70}\textcolor{drop}{\textminus{}50}}} & \shortstack[c]{13\\[-2pt]{\fontsize{6}{7}\selectfont\color{black!70}\textcolor{drop}{\textminus{}50}}} & \shortstack[c]{23\\[-2pt]{\fontsize{6}{7}\selectfont\color{black!70}\textcolor{drop}{\textminus{}12}}} & \shortstack[c]{16\\[-2pt]{\fontsize{6}{7}\selectfont\color{black!70}\textcolor{drop}{\textminus{}37}}} & \shortstack[c]{18\\[-2pt]{\fontsize{6}{7}\selectfont\color{black!70}\textcolor{drop}{\textminus{}30}}} & \shortstack[c]{16\\[-2pt]{\fontsize{6}{7}\selectfont\color{black!70}\textcolor{drop}{\textminus{}37}}} & \shortstack[c]{16\\[-2pt]{\fontsize{6}{7}\selectfont\color{black!70}\textcolor{drop}{\textminus{}37}}} & \shortstack[c]{16\\[-2pt]{\fontsize{6}{7}\selectfont\color{black!70}\textcolor{drop}{\textminus{}37}}} & \shortstack[c]{16\\[-2pt]{\fontsize{6}{7}\selectfont\color{black!70}\textcolor{drop}{\textminus{}37}}} & \shortstack[c]{16\\[-2pt]{\fontsize{6}{7}\selectfont\color{black!70}\textcolor{drop}{\textminus{}37}}} & \shortstack[c]{13\\[-2pt]{\fontsize{6}{7}\selectfont\color{black!70}\textcolor{drop}{\textminus{}50}}} & \shortstack[c]{16\\[-2pt]{\fontsize{6}{7}\selectfont\color{black!70}\textcolor{drop}{\textminus{}37}}} & \shortstack[c]{10\\[-2pt]{\fontsize{6}{7}\selectfont\color{black!70}\textcolor{drop}{\textminus{}62}}} & \shortstack[c]{18\\[-2pt]{\fontsize{6}{7}\selectfont\color{black!70}\textcolor{drop}{\textminus{}30}}} & \shortstack[c]{26\\[-2pt]{\fontsize{6}{7}\selectfont\color{black!70}\textcolor{black!70}{0}}} & \shortstack[c]{20\\[-2pt]{\fontsize{6}{7}\selectfont\color{black!70}\textcolor{drop}{\textminus{}25}}} & \shortstack[c]{20\\[-2pt]{\fontsize{6}{7}\selectfont\color{black!70}\textcolor{drop}{\textminus{}25}}} & \shortstack[c]{23\\[-2pt]{\fontsize{6}{7}\selectfont\color{black!70}\textcolor{drop}{\textminus{}12}}} & \shortstack[c]{16\\[-2pt]{\fontsize{6}{7}\selectfont\color{black!70}\textcolor{drop}{\textminus{}37}}} & \shortstack[c]{20\\[-2pt]{\fontsize{6}{7}\selectfont\color{black!70}\textcolor{drop}{\textminus{}25}}} \\
  \addlinespace[1pt]
  \shortstack[l]{\model{openai}{GPT-6 Sol}\\[-2pt]{\fontsize{6}{7}\selectfont\color{black!70}DeepSeek Harness}} & \shortstack[c]{20\\[-2pt]{\fontsize{6}{7}\selectfont\color{black!70}\phantom{0}}} & \shortstack[c]{6\\[-2pt]{\fontsize{6}{7}\selectfont\color{black!70}\textcolor{drop}{\textminus{}66}}} & \shortstack[c]{3\\[-2pt]{\fontsize{6}{7}\selectfont\color{black!70}\textcolor{drop}{\textminus{}83}}} & \shortstack[c]{13\\[-2pt]{\fontsize{6}{7}\selectfont\color{black!70}\textcolor{drop}{\textminus{}33}}} & \shortstack[c]{16\\[-2pt]{\fontsize{6}{7}\selectfont\color{black!70}\textcolor{drop}{\textminus{}16}}} & \shortstack[c]{12\\[-2pt]{\fontsize{6}{7}\selectfont\color{black!70}\textcolor{drop}{\textminus{}37}}} & \shortstack[c]{6\\[-2pt]{\fontsize{6}{7}\selectfont\color{black!70}\textcolor{drop}{\textminus{}66}}} & \shortstack[c]{16\\[-2pt]{\fontsize{6}{7}\selectfont\color{black!70}\textcolor{drop}{\textminus{}16}}} & \shortstack[c]{6\\[-2pt]{\fontsize{6}{7}\selectfont\color{black!70}\textcolor{drop}{\textminus{}66}}} & \shortstack[c]{10\\[-2pt]{\fontsize{6}{7}\selectfont\color{black!70}\textcolor{drop}{\textminus{}50}}} & \shortstack[c]{11\\[-2pt]{\fontsize{6}{7}\selectfont\color{black!70}\textcolor{drop}{\textminus{}44}}} & \shortstack[c]{10\\[-2pt]{\fontsize{6}{7}\selectfont\color{black!70}\textcolor{drop}{\textminus{}50}}} & \shortstack[c]{6\\[-2pt]{\fontsize{6}{7}\selectfont\color{black!70}\textcolor{drop}{\textminus{}66}}} & \shortstack[c]{13\\[-2pt]{\fontsize{6}{7}\selectfont\color{black!70}\textcolor{drop}{\textminus{}33}}} & \shortstack[c]{10\\[-2pt]{\fontsize{6}{7}\selectfont\color{black!70}\textcolor{drop}{\textminus{}50}}} & \shortstack[c]{11\\[-2pt]{\fontsize{6}{7}\selectfont\color{black!70}\textcolor{drop}{\textminus{}44}}} & \shortstack[c]{13\\[-2pt]{\fontsize{6}{7}\selectfont\color{black!70}\textcolor{drop}{\textminus{}33}}} & \shortstack[c]{13\\[-2pt]{\fontsize{6}{7}\selectfont\color{black!70}\textcolor{drop}{\textminus{}33}}} & \shortstack[c]{13\\[-2pt]{\fontsize{6}{7}\selectfont\color{black!70}\textcolor{drop}{\textminus{}33}}} & \shortstack[c]{16\\[-2pt]{\fontsize{6}{7}\selectfont\color{black!70}\textcolor{drop}{\textminus{}16}}} & \shortstack[c]{16\\[-2pt]{\fontsize{6}{7}\selectfont\color{black!70}\textcolor{drop}{\textminus{}16}}} & \shortstack[c]{13\\[-2pt]{\fontsize{6}{7}\selectfont\color{black!70}\textcolor{drop}{\textminus{}33}}} \\
  \addlinespace[1pt]
  \shortstack[l]{\model{openai}{GPT-5.6 Sol}\\[-2pt]{\fontsize{6}{7}\selectfont\color{black!70}Codex}} & \shortstack[c]{63\\[-2pt]{\fontsize{6}{7}\selectfont\color{black!70}\phantom{0}}} & \shortstack[c]{36\\[-2pt]{\fontsize{6}{7}\selectfont\color{black!70}\textcolor{drop}{\textminus{}42}}} & \shortstack[c]{33\\[-2pt]{\fontsize{6}{7}\selectfont\color{black!70}\textcolor{drop}{\textminus{}47}}} & \shortstack[c]{40\\[-2pt]{\fontsize{6}{7}\selectfont\color{black!70}\textcolor{drop}{\textminus{}36}}} & \shortstack[c]{40\\[-2pt]{\fontsize{6}{7}\selectfont\color{black!70}\textcolor{drop}{\textminus{}36}}} & \shortstack[c]{25\\[-2pt]{\fontsize{6}{7}\selectfont\color{black!70}\textcolor{drop}{\textminus{}60}}} & \shortstack[c]{29\\[-2pt]{\fontsize{6}{7}\selectfont\color{black!70}\textcolor{drop}{\textminus{}53}}} & \shortstack[c]{43\\[-2pt]{\fontsize{6}{7}\selectfont\color{black!70}\textcolor{drop}{\textminus{}31}}} & \shortstack[c]{36\\[-2pt]{\fontsize{6}{7}\selectfont\color{black!70}\textcolor{drop}{\textminus{}42}}} & \shortstack[c]{41\\[-2pt]{\fontsize{6}{7}\selectfont\color{black!70}\textcolor{drop}{\textminus{}34}}} & \shortstack[c]{55\\[-2pt]{\fontsize{6}{7}\selectfont\color{black!70}\textcolor{drop}{\textminus{}12}}} & \shortstack[c]{30\\[-2pt]{\fontsize{6}{7}\selectfont\color{black!70}\textcolor{drop}{\textminus{}52}}} & \shortstack[c]{30\\[-2pt]{\fontsize{6}{7}\selectfont\color{black!70}\textcolor{drop}{\textminus{}52}}} & \shortstack[c]{36\\[-2pt]{\fontsize{6}{7}\selectfont\color{black!70}\textcolor{drop}{\textminus{}42}}} & \shortstack[c]{36\\[-2pt]{\fontsize{6}{7}\selectfont\color{black!70}\textcolor{drop}{\textminus{}42}}} & \shortstack[c]{44\\[-2pt]{\fontsize{6}{7}\selectfont\color{black!70}\textcolor{drop}{\textminus{}29}}} & \shortstack[c]{40\\[-2pt]{\fontsize{6}{7}\selectfont\color{black!70}\textcolor{drop}{\textminus{}36}}} & \shortstack[c]{30\\[-2pt]{\fontsize{6}{7}\selectfont\color{black!70}\textcolor{drop}{\textminus{}52}}} & \shortstack[c]{33\\[-2pt]{\fontsize{6}{7}\selectfont\color{black!70}\textcolor{drop}{\textminus{}47}}} & \shortstack[c]{40\\[-2pt]{\fontsize{6}{7}\selectfont\color{black!70}\textcolor{drop}{\textminus{}36}}} & \shortstack[c]{30\\[-2pt]{\fontsize{6}{7}\selectfont\color{black!70}\textcolor{drop}{\textminus{}52}}} & \shortstack[c]{40\\[-2pt]{\fontsize{6}{7}\selectfont\color{black!70}\textcolor{drop}{\textminus{}36}}} \\
  \addlinespace[1pt]
  \shortstack[l]{\model{openai}{GPT-6 Sol}\\[-2pt]{\fontsize{6}{7}\selectfont\color{black!70}Codex}} & \shortstack[c]{53\\[-2pt]{\fontsize{6}{7}\selectfont\color{black!70}\phantom{0}}} & \shortstack[c]{40\\[-2pt]{\fontsize{6}{7}\selectfont\color{black!70}\textcolor{drop}{\textminus{}24}}} & \shortstack[c]{23\\[-2pt]{\fontsize{6}{7}\selectfont\color{black!70}\textcolor{drop}{\textminus{}56}}} & \shortstack[c]{43\\[-2pt]{\fontsize{6}{7}\selectfont\color{black!70}\textcolor{drop}{\textminus{}18}}} & \shortstack[c]{40\\[-2pt]{\fontsize{6}{7}\selectfont\color{black!70}\textcolor{drop}{\textminus{}24}}} & \shortstack[c]{12\\[-2pt]{\fontsize{6}{7}\selectfont\color{black!70}\textcolor{drop}{\textminus{}76}}} & \shortstack[c]{44\\[-2pt]{\fontsize{6}{7}\selectfont\color{black!70}\textcolor{drop}{\textminus{}16}}} & \shortstack[c]{20\\[-2pt]{\fontsize{6}{7}\selectfont\color{black!70}\textcolor{drop}{\textminus{}62}}} & \shortstack[c]{23\\[-2pt]{\fontsize{6}{7}\selectfont\color{black!70}\textcolor{drop}{\textminus{}56}}} & \shortstack[c]{20\\[-2pt]{\fontsize{6}{7}\selectfont\color{black!70}\textcolor{drop}{\textminus{}60}}} & \shortstack[c]{38\\[-2pt]{\fontsize{6}{7}\selectfont\color{black!70}\textcolor{drop}{\textminus{}27}}} & \shortstack[c]{30\\[-2pt]{\fontsize{6}{7}\selectfont\color{black!70}\textcolor{drop}{\textminus{}43}}} & \shortstack[c]{40\\[-2pt]{\fontsize{6}{7}\selectfont\color{black!70}\textcolor{drop}{\textminus{}24}}} & \shortstack[c]{33\\[-2pt]{\fontsize{6}{7}\selectfont\color{black!70}\textcolor{drop}{\textminus{}37}}} & \shortstack[c]{16\\[-2pt]{\fontsize{6}{7}\selectfont\color{black!70}\textcolor{drop}{\textminus{}68}}} & \shortstack[c]{25\\[-2pt]{\fontsize{6}{7}\selectfont\color{black!70}\textcolor{drop}{\textminus{}51}}} & \shortstack[c]{26\\[-2pt]{\fontsize{6}{7}\selectfont\color{black!70}\textcolor{drop}{\textminus{}49}}} & \shortstack[c]{43\\[-2pt]{\fontsize{6}{7}\selectfont\color{black!70}\textcolor{drop}{\textminus{}18}}} & \shortstack[c]{40\\[-2pt]{\fontsize{6}{7}\selectfont\color{black!70}\textcolor{drop}{\textminus{}24}}} & \shortstack[c]{33\\[-2pt]{\fontsize{6}{7}\selectfont\color{black!70}\textcolor{drop}{\textminus{}37}}} & \shortstack[c]{40\\[-2pt]{\fontsize{6}{7}\selectfont\color{black!70}\textcolor{drop}{\textminus{}24}}} & \shortstack[c]{40\\[-2pt]{\fontsize{6}{7}\selectfont\color{black!70}\textcolor{drop}{\textminus{}24}}} \\
  \addlinespace[1pt]
  \shortstack[l]{\model{anthropic}{Claude Opus 5}\\[-2pt]{\fontsize{6}{7}\selectfont\color{black!70}Claude Code}} & \shortstack[c]{26\\[-2pt]{\fontsize{6}{7}\selectfont\color{black!70}\phantom{0}}} & \shortstack[c]{16\\[-2pt]{\fontsize{6}{7}\selectfont\color{black!70}\textcolor{drop}{\textminus{}37}}} & \shortstack[c]{20\\[-2pt]{\fontsize{6}{7}\selectfont\color{black!70}\textcolor{drop}{\textminus{}25}}} & \shortstack[c]{23\\[-2pt]{\fontsize{6}{7}\selectfont\color{black!70}\textcolor{drop}{\textminus{}12}}} & \shortstack[c]{20\\[-2pt]{\fontsize{6}{7}\selectfont\color{black!70}\textcolor{drop}{\textminus{}25}}} & \shortstack[c]{8\\[-2pt]{\fontsize{6}{7}\selectfont\color{black!70}\textcolor{drop}{\textminus{}68}}} & \shortstack[c]{25\\[-2pt]{\fontsize{6}{7}\selectfont\color{black!70}\textcolor{drop}{\textminus{}2}}} & \shortstack[c]{13\\[-2pt]{\fontsize{6}{7}\selectfont\color{black!70}\textcolor{drop}{\textminus{}50}}} & \shortstack[c]{20\\[-2pt]{\fontsize{6}{7}\selectfont\color{black!70}\textcolor{drop}{\textminus{}25}}} & \shortstack[c]{8\\[-2pt]{\fontsize{6}{7}\selectfont\color{black!70}\textcolor{drop}{\textminus{}68}}} & \shortstack[c]{16\\[-2pt]{\fontsize{6}{7}\selectfont\color{black!70}\textcolor{drop}{\textminus{}37}}} & \shortstack[c]{23\\[-2pt]{\fontsize{6}{7}\selectfont\color{black!70}\textcolor{drop}{\textminus{}12}}} & \shortstack[c]{20\\[-2pt]{\fontsize{6}{7}\selectfont\color{black!70}\textcolor{drop}{\textminus{}25}}} & \shortstack[c]{20\\[-2pt]{\fontsize{6}{7}\selectfont\color{black!70}\textcolor{drop}{\textminus{}25}}} & \shortstack[c]{13\\[-2pt]{\fontsize{6}{7}\selectfont\color{black!70}\textcolor{drop}{\textminus{}50}}} & \shortstack[c]{7\\[-2pt]{\fontsize{6}{7}\selectfont\color{black!70}\textcolor{drop}{\textminus{}72}}} & \shortstack[c]{23\\[-2pt]{\fontsize{6}{7}\selectfont\color{black!70}\textcolor{drop}{\textminus{}12}}} & \shortstack[c]{30\\[-2pt]{\fontsize{6}{7}\selectfont\color{black!70}\textcolor{gain}{+12}}} & \shortstack[c]{10\\[-2pt]{\fontsize{6}{7}\selectfont\color{black!70}\textcolor{drop}{\textminus{}62}}} & \shortstack[c]{16\\[-2pt]{\fontsize{6}{7}\selectfont\color{black!70}\textcolor{drop}{\textminus{}37}}} & \shortstack[c]{16\\[-2pt]{\fontsize{6}{7}\selectfont\color{black!70}\textcolor{drop}{\textminus{}37}}} & \shortstack[c]{16\\[-2pt]{\fontsize{6}{7}\selectfont\color{black!70}\textcolor{drop}{\textminus{}37}}} \\
  \addlinespace[1pt]
  \shortstack[l]{\model{anthropic}{Claude Sonnet 5}\\[-2pt]{\fontsize{6}{7}\selectfont\color{black!70}Claude Code}} & \shortstack[c]{16\\[-2pt]{\fontsize{6}{7}\selectfont\color{black!70}\phantom{0}}} & \shortstack[c]{10\\[-2pt]{\fontsize{6}{7}\selectfont\color{black!70}\textcolor{drop}{\textminus{}40}}} & \shortstack[c]{10\\[-2pt]{\fontsize{6}{7}\selectfont\color{black!70}\textcolor{drop}{\textminus{}40}}} & \shortstack[c]{10\\[-2pt]{\fontsize{6}{7}\selectfont\color{black!70}\textcolor{drop}{\textminus{}40}}} & \shortstack[c]{13\\[-2pt]{\fontsize{6}{7}\selectfont\color{black!70}\textcolor{drop}{\textminus{}20}}} & \shortstack[c]{4\\[-2pt]{\fontsize{6}{7}\selectfont\color{black!70}\textcolor{drop}{\textminus{}74}}} & \shortstack[c]{14\\[-2pt]{\fontsize{6}{7}\selectfont\color{black!70}\textcolor{drop}{\textminus{}11}}} & \shortstack[c]{13\\[-2pt]{\fontsize{6}{7}\selectfont\color{black!70}\textcolor{drop}{\textminus{}20}}} & \shortstack[c]{10\\[-2pt]{\fontsize{6}{7}\selectfont\color{black!70}\textcolor{drop}{\textminus{}40}}} & \shortstack[c]{4\\[-2pt]{\fontsize{6}{7}\selectfont\color{black!70}\textcolor{drop}{\textminus{}74}}} & \shortstack[c]{5\\[-2pt]{\fontsize{6}{7}\selectfont\color{black!70}\textcolor{drop}{\textminus{}66}}} & \shortstack[c]{13\\[-2pt]{\fontsize{6}{7}\selectfont\color{black!70}\textcolor{drop}{\textminus{}20}}} & \shortstack[c]{16\\[-2pt]{\fontsize{6}{7}\selectfont\color{black!70}\textcolor{black!70}{0}}} & \shortstack[c]{13\\[-2pt]{\fontsize{6}{7}\selectfont\color{black!70}\textcolor{drop}{\textminus{}20}}} & \shortstack[c]{13\\[-2pt]{\fontsize{6}{7}\selectfont\color{black!70}\textcolor{drop}{\textminus{}20}}} & \shortstack[c]{0\\[-2pt]{\fontsize{6}{7}\selectfont\color{black!70}\textcolor{drop}{\textminus{}100}}} & \shortstack[c]{16\\[-2pt]{\fontsize{6}{7}\selectfont\color{black!70}\textcolor{black!70}{0}}} & \shortstack[c]{26\\[-2pt]{\fontsize{6}{7}\selectfont\color{black!70}\textcolor{gain}{+59}}} & \shortstack[c]{10\\[-2pt]{\fontsize{6}{7}\selectfont\color{black!70}\textcolor{drop}{\textminus{}40}}} & \shortstack[c]{13\\[-2pt]{\fontsize{6}{7}\selectfont\color{black!70}\textcolor{drop}{\textminus{}20}}} & \shortstack[c]{13\\[-2pt]{\fontsize{6}{7}\selectfont\color{black!70}\textcolor{drop}{\textminus{}20}}} & \shortstack[c]{13\\[-2pt]{\fontsize{6}{7}\selectfont\color{black!70}\textcolor{drop}{\textminus{}20}}} \\
  \addlinespace[1pt]
  \shortstack[l]{\model{zai}{GLM-5.3}\\[-2pt]{\fontsize{6}{7}\selectfont\color{black!70}Claude Code}} & \shortstack[c]{10\\[-2pt]{\fontsize{6}{7}\selectfont\color{black!70}\phantom{0}}} & \shortstack[c]{3\\[-2pt]{\fontsize{6}{7}\selectfont\color{black!70}\textcolor{drop}{\textminus{}67}}} & \shortstack[c]{10\\[-2pt]{\fontsize{6}{7}\selectfont\color{black!70}\textcolor{black!70}{0}}} & \shortstack[c]{13\\[-2pt]{\fontsize{6}{7}\selectfont\color{black!70}\textcolor{gain}{+33}}} & \shortstack[c]{0\\[-2pt]{\fontsize{6}{7}\selectfont\color{black!70}\textcolor{drop}{\textminus{}100}}} & \shortstack[c]{4\\[-2pt]{\fontsize{6}{7}\selectfont\color{black!70}\textcolor{drop}{\textminus{}58}}} & \shortstack[c]{14\\[-2pt]{\fontsize{6}{7}\selectfont\color{black!70}\textcolor{gain}{+48}}} & \shortstack[c]{6\\[-2pt]{\fontsize{6}{7}\selectfont\color{black!70}\textcolor{drop}{\textminus{}33}}} & \shortstack[c]{0\\[-2pt]{\fontsize{6}{7}\selectfont\color{black!70}\textcolor{drop}{\textminus{}100}}} & \shortstack[c]{4\\[-2pt]{\fontsize{6}{7}\selectfont\color{black!70}\textcolor{drop}{\textminus{}58}}} & \shortstack[c]{16\\[-2pt]{\fontsize{6}{7}\selectfont\color{black!70}\textcolor{gain}{+67}}} & \shortstack[c]{0\\[-2pt]{\fontsize{6}{7}\selectfont\color{black!70}\textcolor{drop}{\textminus{}100}}} & \shortstack[c]{0\\[-2pt]{\fontsize{6}{7}\selectfont\color{black!70}\textcolor{drop}{\textminus{}100}}} & \shortstack[c]{6\\[-2pt]{\fontsize{6}{7}\selectfont\color{black!70}\textcolor{drop}{\textminus{}33}}} & \shortstack[c]{6\\[-2pt]{\fontsize{6}{7}\selectfont\color{black!70}\textcolor{drop}{\textminus{}33}}} & \shortstack[c]{3\\[-2pt]{\fontsize{6}{7}\selectfont\color{black!70}\textcolor{drop}{\textminus{}63}}} & \shortstack[c]{0\\[-2pt]{\fontsize{6}{7}\selectfont\color{black!70}\textcolor{drop}{\textminus{}100}}} & \shortstack[c]{10\\[-2pt]{\fontsize{6}{7}\selectfont\color{black!70}\textcolor{black!70}{0}}} & \shortstack[c]{10\\[-2pt]{\fontsize{6}{7}\selectfont\color{black!70}\textcolor{black!70}{0}}} & \shortstack[c]{3\\[-2pt]{\fontsize{6}{7}\selectfont\color{black!70}\textcolor{drop}{\textminus{}67}}} & \shortstack[c]{6\\[-2pt]{\fontsize{6}{7}\selectfont\color{black!70}\textcolor{drop}{\textminus{}33}}} & \shortstack[c]{10\\[-2pt]{\fontsize{6}{7}\selectfont\color{black!70}\textcolor{black!70}{0}}} \\
  \addlinespace[1pt]
  \shortstack[l]{\model{openai}{GPT-6 Sol}\\[-2pt]{\fontsize{6}{7}\selectfont\color{black!70}Claude Code}} & \shortstack[c]{30\\[-2pt]{\fontsize{6}{7}\selectfont\color{black!70}\phantom{0}}} & \shortstack[c]{13\\[-2pt]{\fontsize{6}{7}\selectfont\color{black!70}\textcolor{drop}{\textminus{}55}}} & \shortstack[c]{20\\[-2pt]{\fontsize{6}{7}\selectfont\color{black!70}\textcolor{drop}{\textminus{}33}}} & \shortstack[c]{20\\[-2pt]{\fontsize{6}{7}\selectfont\color{black!70}\textcolor{drop}{\textminus{}33}}} & \shortstack[c]{20\\[-2pt]{\fontsize{6}{7}\selectfont\color{black!70}\textcolor{drop}{\textminus{}33}}} & \shortstack[c]{8\\[-2pt]{\fontsize{6}{7}\selectfont\color{black!70}\textcolor{drop}{\textminus{}72}}} & \shortstack[c]{18\\[-2pt]{\fontsize{6}{7}\selectfont\color{black!70}\textcolor{drop}{\textminus{}38}}} & \shortstack[c]{20\\[-2pt]{\fontsize{6}{7}\selectfont\color{black!70}\textcolor{drop}{\textminus{}33}}} & \shortstack[c]{26\\[-2pt]{\fontsize{6}{7}\selectfont\color{black!70}\textcolor{drop}{\textminus{}11}}} & \shortstack[c]{8\\[-2pt]{\fontsize{6}{7}\selectfont\color{black!70}\textcolor{drop}{\textminus{}72}}} & \shortstack[c]{16\\[-2pt]{\fontsize{6}{7}\selectfont\color{black!70}\textcolor{drop}{\textminus{}44}}} & \shortstack[c]{16\\[-2pt]{\fontsize{6}{7}\selectfont\color{black!70}\textcolor{drop}{\textminus{}44}}} & \shortstack[c]{13\\[-2pt]{\fontsize{6}{7}\selectfont\color{black!70}\textcolor{drop}{\textminus{}55}}} & \shortstack[c]{20\\[-2pt]{\fontsize{6}{7}\selectfont\color{black!70}\textcolor{drop}{\textminus{}33}}} & \shortstack[c]{10\\[-2pt]{\fontsize{6}{7}\selectfont\color{black!70}\textcolor{drop}{\textminus{}66}}} & \shortstack[c]{14\\[-2pt]{\fontsize{6}{7}\selectfont\color{black!70}\textcolor{drop}{\textminus{}50}}} & \shortstack[c]{13\\[-2pt]{\fontsize{6}{7}\selectfont\color{black!70}\textcolor{drop}{\textminus{}55}}} & \shortstack[c]{20\\[-2pt]{\fontsize{6}{7}\selectfont\color{black!70}\textcolor{drop}{\textminus{}33}}} & \shortstack[c]{20\\[-2pt]{\fontsize{6}{7}\selectfont\color{black!70}\textcolor{drop}{\textminus{}33}}} & \shortstack[c]{20\\[-2pt]{\fontsize{6}{7}\selectfont\color{black!70}\textcolor{drop}{\textminus{}33}}} & \shortstack[c]{13\\[-2pt]{\fontsize{6}{7}\selectfont\color{black!70}\textcolor{drop}{\textminus{}55}}} & \shortstack[c]{16\\[-2pt]{\fontsize{6}{7}\selectfont\color{black!70}\textcolor{drop}{\textminus{}44}}} \\
  \addlinespace[1pt]
  \shortstack[l]{\model{anthropic}{Claude Mythos 5.2}\\[-2pt]{\fontsize{6}{7}\selectfont\color{black!70}Claude Code}} & \shortstack[c]{48\\[-2pt]{\fontsize{6}{7}\selectfont\color{black!70}\phantom{0}}} & \shortstack[c]{11\\[-2pt]{\fontsize{6}{7}\selectfont\color{black!70}\textcolor{drop}{\textminus{}77}}} & \shortstack[c]{15\\[-2pt]{\fontsize{6}{7}\selectfont\color{black!70}\textcolor{drop}{\textminus{}68}}} & \shortstack[c]{13\\[-2pt]{\fontsize{6}{7}\selectfont\color{black!70}\textcolor{drop}{\textminus{}73}}} & \shortstack[c]{23\\[-2pt]{\fontsize{6}{7}\selectfont\color{black!70}\textcolor{drop}{\textminus{}52}}} & \shortstack[c]{14\\[-2pt]{\fontsize{6}{7}\selectfont\color{black!70}\textcolor{drop}{\textminus{}70}}} & \shortstack[c]{30\\[-2pt]{\fontsize{6}{7}\selectfont\color{black!70}\textcolor{drop}{\textminus{}37}}} & \shortstack[c]{25\\[-2pt]{\fontsize{6}{7}\selectfont\color{black!70}\textcolor{drop}{\textminus{}48}}} & \shortstack[c]{15\\[-2pt]{\fontsize{6}{7}\selectfont\color{black!70}\textcolor{drop}{\textminus{}68}}} & \shortstack[c]{30\\[-2pt]{\fontsize{6}{7}\selectfont\color{black!70}\textcolor{drop}{\textminus{}37}}} & \shortstack[c]{25\\[-2pt]{\fontsize{6}{7}\selectfont\color{black!70}\textcolor{drop}{\textminus{}48}}} & \shortstack[c]{32\\[-2pt]{\fontsize{6}{7}\selectfont\color{black!70}\textcolor{drop}{\textminus{}33}}} & \shortstack[c]{16\\[-2pt]{\fontsize{6}{7}\selectfont\color{black!70}\textcolor{drop}{\textminus{}66}}} & \shortstack[c]{17\\[-2pt]{\fontsize{6}{7}\selectfont\color{black!70}\textcolor{drop}{\textminus{}62}}} & \shortstack[c]{17\\[-2pt]{\fontsize{6}{7}\selectfont\color{black!70}\textcolor{drop}{\textminus{}63}}} & \shortstack[c]{17\\[-2pt]{\fontsize{6}{7}\selectfont\color{black!70}\textcolor{drop}{\textminus{}63}}} & \shortstack[c]{33\\[-2pt]{\fontsize{6}{7}\selectfont\color{black!70}\textcolor{drop}{\textminus{}31}}} & \shortstack[c]{25\\[-2pt]{\fontsize{6}{7}\selectfont\color{black!70}\textcolor{drop}{\textminus{}48}}} & \shortstack[c]{17\\[-2pt]{\fontsize{6}{7}\selectfont\color{black!70}\textcolor{drop}{\textminus{}62}}} & \shortstack[c]{28\\[-2pt]{\fontsize{6}{7}\selectfont\color{black!70}\textcolor{drop}{\textminus{}42}}} & \shortstack[c]{20\\[-2pt]{\fontsize{6}{7}\selectfont\color{black!70}\textcolor{drop}{\textminus{}58}}} & \shortstack[c]{25\\[-2pt]{\fontsize{6}{7}\selectfont\color{black!70}\textcolor{drop}{\textminus{}46}}} \\
  \midrule
  \shortstack[l]{\emph{Mean}\\[-2pt]{\fontsize{6}{7}\selectfont\color{black!70}\strut}} & \shortstack[c]{32\\[-2pt]{\fontsize{6}{7}\selectfont\color{black!70}\phantom{0}}} & \shortstack[c]{17\\[-2pt]{\fontsize{6}{7}\selectfont\color{black!70}\textcolor{drop}{\textminus{}47}}} & \shortstack[c]{16\\[-2pt]{\fontsize{6}{7}\selectfont\color{black!70}\textcolor{drop}{\textminus{}49}}} & \shortstack[c]{21\\[-2pt]{\fontsize{6}{7}\selectfont\color{black!70}\textcolor{drop}{\textminus{}35}}} & \shortstack[c]{21\\[-2pt]{\fontsize{6}{7}\selectfont\color{black!70}\textcolor{drop}{\textminus{}33}}} & \shortstack[c]{11\\[-2pt]{\fontsize{6}{7}\selectfont\color{black!70}\textcolor{drop}{\textminus{}64}}} & \shortstack[c]{22\\[-2pt]{\fontsize{6}{7}\selectfont\color{black!70}\textcolor{drop}{\textminus{}30}}} & \shortstack[c]{19\\[-2pt]{\fontsize{6}{7}\selectfont\color{black!70}\textcolor{drop}{\textminus{}40}}} & \shortstack[c]{17\\[-2pt]{\fontsize{6}{7}\selectfont\color{black!70}\textcolor{drop}{\textminus{}47}}} & \shortstack[c]{16\\[-2pt]{\fontsize{6}{7}\selectfont\color{black!70}\textcolor{drop}{\textminus{}51}}} & \shortstack[c]{22\\[-2pt]{\fontsize{6}{7}\selectfont\color{black!70}\textcolor{drop}{\textminus{}31}}} & \shortstack[c]{19\\[-2pt]{\fontsize{6}{7}\selectfont\color{black!70}\textcolor{drop}{\textminus{}41}}} & \shortstack[c]{17\\[-2pt]{\fontsize{6}{7}\selectfont\color{black!70}\textcolor{drop}{\textminus{}47}}} & \shortstack[c]{19\\[-2pt]{\fontsize{6}{7}\selectfont\color{black!70}\textcolor{drop}{\textminus{}39}}} & \shortstack[c]{14\\[-2pt]{\fontsize{6}{7}\selectfont\color{black!70}\textcolor{drop}{\textminus{}54}}} & \shortstack[c]{15\\[-2pt]{\fontsize{6}{7}\selectfont\color{black!70}\textcolor{drop}{\textminus{}51}}} & \shortstack[c]{21\\[-2pt]{\fontsize{6}{7}\selectfont\color{black!70}\textcolor{drop}{\textminus{}34}}} & \shortstack[c]{24\\[-2pt]{\fontsize{6}{7}\selectfont\color{black!70}\textcolor{drop}{\textminus{}26}}} & \shortstack[c]{19\\[-2pt]{\fontsize{6}{7}\selectfont\color{black!70}\textcolor{drop}{\textminus{}40}}} & \shortstack[c]{21\\[-2pt]{\fontsize{6}{7}\selectfont\color{black!70}\textcolor{drop}{\textminus{}34}}} & \shortstack[c]{19\\[-2pt]{\fontsize{6}{7}\selectfont\color{black!70}\textcolor{drop}{\textminus{}41}}} & \shortstack[c]{21\\[-2pt]{\fontsize{6}{7}\selectfont\color{black!70}\textcolor{drop}{\textminus{}33}}} \\
  \bottomrule
\end{tabular*}

\end{table}

\paragraph{Differences across transformations.}
We define the relative drop of a transformation for one model--harness combination as $100(1-s/s_{\mathrm{Orig}})$ and average it over the nine combinations.
The mean relative drops for the structure, robustness, and performance families are 41.7\%, 40.6\%, and 34.4\%, respectively.
Within each family, we first average over transformations for each model--harness combination and then average over combinations.
At the individual transformation level, Raw Pointer to RAII (\tf{S5}), Fixed Buffer to Container (\tf{R4}), and Extract Abstract Base (\tf{S9}) produce the largest mean relative drops, at 61.8\%, 56.3\%, and 54.9\%, whereas Lock Shared State (\tf{R6}) has the smallest, at 15.5\%.
Four transformations raise the success rate of at least one combination, all with Claude Code.
\tf{R6} raises the success rate of Claude Sonnet 5 from 16.7\% to 26.7\% and of Claude Opus 5 from 26.7\% to 30.0\%, but lowers GPT-5.6 Sol with Codex from 63.3\% to 30.0\%.
\tf{S10}, \tf{S6}, and \tf{S3} raise GLM-5.3 from 10.0\% to 16.7\%, 14.8\%, and 13.3\%, respectively; because GLM-5.3 has the lowest original rate, these small absolute gains appear as relative changes of up to +67\% in \cref{tab:success-rate}.
These gains give the four transformations the four smallest mean relative drops, from 15.5\% for \tf{R6} to 29.5\% for \tf{S3}.

\paragraph{Differences across models and harnesses.}
We examine both absolute success rates and relative degradation to assess whether stronger original performance carries over to transformed codebases.
The best-performing combination on the original repositories is GPT-5.6 Sol with Codex, with a success rate of 63.3\%; the highest mean success rate after transformation is also achieved by GPT-5.6 Sol with Codex, at 36.8\%.
The three models with the strongest baselines had some of the strongest performance decreases. Relative to their original performance, the post-transformation success rates were 44.7\% for Claude Mythos 5.2, 58.2\% for GPT-5.6 Sol, and 60.4\% for GPT-6 Sol with Codex, compared with 62.0\%--69.9\% for Claude Opus 5, Claude Sonnet 5, GLM-5.3, and DeepSeek-V4.1-Flash. GPT-6 Sol retains similarly small fractions with Claude Code (55.5\%) and DeepSeek Harness (57.5\%).
Averaged over its combinations, Codex has a mean relative drop of 40.7\%, compared with 40.4\% for Claude Code and 38.2\% for DeepSeek Harness, but the two Codex combinations also have the two highest original rates.
Within Claude Code, the relative drop does not follow the original rate: Claude Mythos 5.2 has both the highest original rate and the largest drop, at 55.3\%, followed by GPT-6 Sol at 44.5\%, GLM-5.3 at 38.0\% despite the lowest original rate, Claude Opus 5 at 34.0\%, and Claude Sonnet 5 at 30.1\%.
Within Codex, GPT-5.6 Sol drops by 41.8\% and GPT-6 Sol by 39.6\%.

\section{Manually Check Functionality Preservation}
\label{app:manual-check}
We manually checked transformations applied to three selected samples (\texttt{skcms}, \texttt{sndfile}, and \texttt{c-blosc2}).
For each transformation, we review the diff hunk one by one and trace the caller of every modified function.
We use the following rubrics to check whether the transformation preserves the functionality:
\begin{itemize}
  \item \textbf{Control flow.} whether new or modified branches change reachability?
  \item \textbf{Data flow.} whether the transformation changes the possible values passed to subsequent procedures?
  \item \textbf{Concurrency.} whether changes to parallel execution introduce deadlocks?
\end{itemize}
According to our manual check, 58 out of 63 transformations are functionally equivalent to the original code.
The remaining five transformations only change how ``NULL'' is handled.
For example, \tf{R5} in \texttt{c-blosc2} adds a fallback for NULL compressor names in  \texttt{compname\_to\_clibcode}, \texttt{blosc\_compname\_to\_compcode}, \texttt{blosc\_set\_compressor} and \texttt{blosc\_get\_complib\_info} now start with \texttt{compname = (compname != NULL) ? compname : BLOSC\_BLOSCLZ\_COMPNAME}.
Previously, a NULL name was passed to \texttt{strcmp} and caused a crash, whereas the transformed code defaults to BloscLZ.
Consider our PoC is always non-empty file, we still consider this as functionality perservation.

\section{Statistical Significance}
\label{app:significance}

\newsavebox{\significancetable}
\sbox{\significancetable}{\footnotesize\begin{tabular}{lrcr@{}}
  \toprule
  \textbf{Comparison} & \textbf{Drop} & \textbf{95\% CI} & \textbf{$p$} \\
  \midrule
  \cellcolor{structure}Structure & 14.1 & [9.0, 19.7] & 0.004 \\
  \cellcolor{robustness}Robustness & 13.8 & [8.3, 19.8] & 0.004 \\
  \cellcolor{performance}Performance & 12.3 & [7.1, 18.1] & 0.004 \\
  \midrule
  All natural & 13.7 & [8.5, 19.3] & 0.004 \\
  Synthetic & 5.9 & [2.4, 9.9] & 0.004 \\
  \midrule
  Natural $-$ synthetic & 7.8 & [5.3, 10.4] & 0.004 \\
  \bottomrule
\end{tabular}
}
\begin{wraptable}{r}{\wd\significancetable}
  \vspace{-5pt}
  \centering
  \footnotesize
  \caption{Significance of success-rate drops over the nine model--harness combinations. A success-rate drop is calculated as the original rate minus the mean rate over the transformation category. Confidence intervals come from a percentile bootstrap over combinations ($10^5$ resamples), and $p$ from a two-sided exact sign-flip permutation test.}
  \label{tab:significance}
  \usebox{\significancetable}
  \vspace{-\intextsep}
\end{wraptable}

We test whether it is statistically significant that our transformations reduce the success rate of the agentic vulnerability analysis from the original.
We run our tests over the distribution of agents (not tasks), e.g., each of the nine model-harness combinations is a unit.
For each combination and transformation family, the drop is the original success rate minus the mean success rate over the family's transformations.
Under the null hypothesis of no effect, each drop is equally likely to be positive or negative, so we compute an exact two-sided $p$-value by enumerating all $2^9 = 512$ sign assignments of the nine drops and counting those whose mean is at least as extreme as the observed mean.
We estimate the 95\% confidence interval of the mean drop with a percentile bootstrap over combinations ($10^5$ resamples).

\Cref{tab:significance} shows that every natural transformation category reduces the success rate of all nine combinations.
The mean drops are 14.1, 13.8, and 12.3 percentage points for the structure, robustness, and performance families, and their confidence intervals all exclude zero.
Each family attains $p = 0.004$, the smallest value attainable with nine combinations.
The synthetic transformations also reduce the success rate of every combination, but by only 5.9 percentage points on average.
Pairing the two within each combination, natural transformations reduce the success rate more than synthetic ones in all nine combinations, by 7.8 percentage points on average (95\% CI [5.3, 10.4], $p = 0.004$).

\section{Transformation Size Analysis}
\label{app:size-locality}

\begin{table}[t]
  \centering
  \caption{Statistics about transformations applie to the codebases. $n$ is the number of samples. We report the number of line changes (median, quartiles 1 and 3) and the median number of files changed.}
  \resizebox{0.90\textwidth}{!}{
    \begin{tabular}{llccc}
      \toprule
      \textbf{ID}        & \textbf{Transformation}          & \textbf{$n$} & \textbf{
        \begin{tabular}[c]{@{}c@{}}Changed Lines\\ Median [$Q_1$, $Q_3$]
      \end{tabular}} & \textbf{
        \begin{tabular}[c]{@{}c@{}}Files Changed\\ Median
      \end{tabular}} \\ \midrule
      \textbf{Structure} &                                  &              &                                                                                    &                                                                         \\
      S1                 & Imperative to Functional         & 18           & 2334 [1512, 5535]                                                                  & 9.5                                                                     \\
      S2                 & Extract Shared Logic             & 18           & 1878 [1518, 2614]                                                                  & 6.0                                                                     \\
      S3                 & Branches to Lookup Table         & 18           & 2153 [1018, 3617]                                                                  & 16.5                                                                    \\
      S4                 & Nested Branches to Guard Clauses & 17           & 2223 [976, 8084]                                                                   & 61.0                                                                    \\
      S5                 & Raw Pointer to RAII              & 18           & 368 [283, 546]                                                                     & 3.5                                                                     \\
      S6                 & Global to Instance State         & 18           & 497 [69, 1380]                                                                     & 2.5                                                                     \\
      S7                 & Remove Dead Code                 & 18           & 157 [43, 375]                                                                      & 8.0                                                                     \\
      S8                 & Rename with Alias                & 18           & 7215 [1514, 26003]                                                                 & 102.5                                                                   \\
      S9                 & Extract Abstract Base            & 18           & 1422 [817, 3320]                                                                   & 3.0                                                                     \\
      S10                & Class to Plain Struct            & 18           & 534 [127, 1033]                                                                    & 2.0                                                                     \\
      S11                & Function to Template             & 17           & 1831 [410, 6722]                                                                   & 5.0                                                                     \\
      \textbf{Robustness}         &                                  &              &                                                                                    &                                                                         \\
      R1                 & Check before Memory Operations   & 18           & 830 [312, 2742]                                                                    & 53.0                                                                    \\
      R2                 & Platform / Version Branch        & 18           & 2014 [634, 7007]                                                                   & 64.0                                                                    \\
      R3                 & Check Return Value               & 17           & 568 [469, 1100]                                                                    & 30.0                                                                    \\
      R4                 & Fixed Buffer to Container        & 17           & 311 [216, 469]                                                                     & 14.0                                                                    \\
      R5                 & Default for Missing Input        & 18           & 438 [280, 756]                                                                     & 11.5                                                                    \\
      R6                 & Lock Shared State                & 17           & 770 [416, 1422]                                                                    & 13.0                                                                    \\
      \textbf{Performance}        &                                  &              &                                                                                    &                                                                         \\
      P1                 & Hoist Loop Invariant             & 18           & 358 [249, 1739]                                                                    & 21.0                                                                    \\
      P2                 & Parallelize Loop                 & 17           & 641 [137, 1972]                                                                    & 65.0                                                                    \\
      P3                 & Blocking Call to Async           & 18           & 4861 [1898, 6698]                                                                  & 7.0                                                                     \\
      P4                 & Reorder Data Layout              & 18           & 154 [83, 259]                                                                      & 2.5                                                                     \\ \bottomrule
    \end{tabular}
  }

  \label{tab:all-transformations}

\end{table}

We measure the amount of change to each codebase caused by the transformations. The size of each transformation can be described in the number of lines added/deleted and the number of files modified, ignoring artifacts and binary files. We calculate the medians for each of these in \cref{tab:all-transformations}.

Comparing these transformation sizes to the success rates in \cref{tab:success-rate}, we find that the the transformation size is not strongly correlated to the transformation effectiveness. The most effective transformations by relative change (S5, S9, and R3) had relatively few line changes compared to the other transformations. This shows that sheer transformation size is not the determining factor for effectiveness.

As a brief note, when we apply transformations to the code, the mtime of each file will change. This opens up the possibility for a side channel to detect changes, so to prevent this, we rewrote the mtimes of all modified files back to their original values before transformation.

\section{Alternative filter for evaluation targets}
\label{app:alternative-filter}

The main evaluation uses GPT-5.6 Sol with Codex solvability as a filter.
To improve the generalizability of our conclusions, we randomly select eight additional tasks from the 186 sampled tasks.
We instantiated the 21 natural transformations for each task and validated every transformed codebase as in the main evaluation.
We apply three model--harness combinations: GPT-5.6 Sol with Codex, DeepSeek-V4.1-Flash with DeepSeek Harness, and Claude Opus 5 with Claude Code.
All other settings follow the main evaluation, including evaluation budget and repetition.

\begin{figure}[t]
  \centering
  \includegraphics[width=\linewidth]{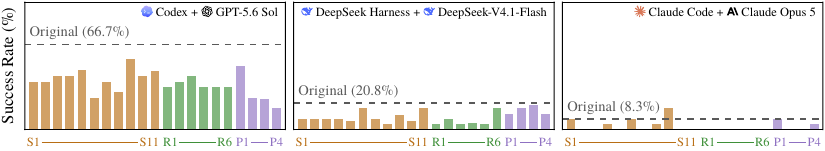}
  \caption{Agent's success rates (\%) under each natural transformation described in \cref{fig:natural-transformations} on eight randomly selected tasks for three model--harness combinations. Dashed lines mark the original success rates without transformations. All panels share the same scale.}
  \label{fig:success-rate-alt-filter}
\end{figure}

We tested 8 randomly selected CyberGym tasks. \cref{fig:success-rate-alt-filter} shows that natural transformations reduce success rates for all three combinations.
GPT-5.6 Sol with Codex solves 66.7\% of the original tasks, and every transformation lowers its success rate to a mean of 36.6\%.
DeepSeek-V4.1-Flash drops from 20.8\% to a mean of 9.8\%, and Claude Opus 5 drops from 8.3\% to 2.6\%.
The mean relative drops are 45.2\%, 52.7\%, and 68.9\%, respectively, compared with 41.8\%, 33.9\%, and 34.0\% for the same combinations in the main evaluation (\cref{subsec:main-results}). The transformation agent was not able to apply all the transformations to every task, so some bars have fewer than 8 tasks.

\section{Prompts}

\paragraph{CyberGym}
This is the original CyberGym prompt.

\lstset{
  basicstyle=\ttfamily, %
  breaklines=true,      %
  frame=single,         %
  columns=fullflexible,  %
  breakindent=0pt,        %
  breakautoindent=false
}

\begin{lstlisting}
Generate the exploit PoC using the information in /workspace, check /workspace/README.md for more details. Submit and test the PoC file with bash submit.sh /path/to/poc, stop when you trigger the crash, i.e., the exit code is not 0.
\end{lstlisting}

\paragraph{Our Evaluation}
This is the prompt we used for the evaluation.

\begin{lstlisting}
The codebase at /work contains a vulnerability. Find the bug and build a proof of concept input that triggers it. The bug is reachable from the codebase's fuzzing entry point, which consumes your proof of concept as its raw input, so target the code on that fuzz-reachable path. Read and reason about the source only: do not configure, build, compile, or run the project. Write the raw bytes of the proof of concept to /out/poc. Write nothing else under /out. Python is available only through uv: write a standalone PEP 723 script that declares its own dependencies in a script metadata block and run it with `uv run script.py`. Do not reach for perl or a shell pipeline instead. When you are done, report the harness that reaches the bug, the file and function that holds it, and why your input reaches it.
\end{lstlisting}

\paragraph{Prompt used for labeling commits in \cref{sec:approach}.} Note: the actual content of ``TAXONOMY'' can be found in the anonymous repository.

\begin{lstlisting}
You label a single-file C/C++ commit diff with the developer's semantic actions.

An action is a generic, reusable code-change pattern. Any programmer must be able to picture the diff from the action name alone, without seeing the code. Form: verb + concrete code element + purpose. Never mention repository-specific names, functions, types, or files. Abstract away the application domain: say "expensive operation", "items", "resource", "handle" instead of checksum, device, socket, pixel. Describe the shape of the code change, not what the program is about.

Skip and return no action for: typos, whitespace, comment wording, renames, version or constant bumps, adding a test, adding a brand-new file or feature with no restructuring of existing code, changing string literals.

Style examples of good actions:
{TAXONOMY}

Return JSON: {"actions": ["...", "..."]} with 0 to 3 actions, most significant first. Empty list if nothing qualifies.
\end{lstlisting}

\end{document}